# Lyα absorption at transits of HD 209458b: comparative study of various mechanisms under different conditions

Khodachenko[1,2] M. L., Shaikhislamov[3] I. F., Lammer[1] H., Kislyakova[1] K. G., Fossati[1] L., Johnstone[4] C. P., Arkhypov[1] O.V., Berezutsky[3] A. G., Miroshnichenko[3] I. B., Posukh[3] V. G.

1) Space Research Institute, Austrian Acad. Sci., Graz, Austria
2) Skobeltsyn Institute of Nuclear Physics, Moscow State University, Moscow, Russia
3) Institute of Laser Physics SB RAS, Novosibirsk, Russia
4) Dep. of Astrophysics, University of Vienna

E-mail address: maxim.khodachenko@oeaw.ac.at

**Abstract:**
To shed more light on the nature of the observed Lyα absorption during transits of HD209458b and to quantify the major mechanisms responsible for the production of fast hydrogen atoms (so-called ENAs) around the planet, a 2D hydrodynamic multi-fluid modelling of the expanding planetary upper atmosphere, driven by the stellar XUV, and its interaction with the stellar wind has been performed. The model self-consistently describes the escaping planetary wind, taking into account the generation of ENAs due to particle acceleration by the radiation pressure and by charge-exchange between the stellar wind protons and planetary atoms. The calculations in a wide range of stellar wind parameters and XUV flux values showed that under the typical Sun-like star conditions the amount of generated ENAs is too small, and the observed absorption at the level of 6÷8% can be attributed only to the non-resonant natural line broadening. For lower XUV fluxes, e.g., during the activity minima, the number of planetary atoms that survive photo-ionization and give the origin to ENAs, increases resulting in up to 10÷15% absorption at blue wing of Lyα line, caused by the resonant thermal line broadening. Similar asymmetric absorption can be seen under the conditions realized during Coronal Mass Ejections, when sufficiently high stellar wind pressure confines the escaping planetary material within a kind of bowshock around the planet. It has been found, that the radiation pressure in all considered cases has a negligible contribution to the production of ENAs and the corresponding absorption.

## 1. Introduction

The study of planets beyond the Solar system, or exoplanets, is one of the fast-growing and intriguing fields in the space science. After just 20 years of the subject history more than 3500 exoplanets, including about 600 multiple systems, have been discovered (e.g., NASA exoplanet archive https://exoplanetarchive.ipac.caltech.edu/). Among those more than 2700 are transiting exoplanets, which admit spectral probing. A large number of Jupiter-type giant exoplanets, and even larger amount of less massive planets are found to orbit at extremely close distances ≤0.05 a.u. (e.g., http://exoplanets.org/). *Lammer et al. (2003)* were the first to show that a hydrogen-rich atmosphere of such planets will be heated to several thousand Kelvin and dynamically expand. *Vidal-Madjar et al. (2003)* observed the transiting exoplanet HD 209458b with the HST/STIS-instrument and discovered a 15% intensity drop in the high velocity blue part of the stellar Lyα line which can be explained only by presence of energetic neutral atoms (ENAs) of planetary origin, moving away from the star with velocity up to 150 km/s. Further re-analysis of data gave slightly less and more symmetric absorption profile at the level of 6-9% (*Ben-Jaffel 2007, Vidal-Madjar et al. 2008; Ben-Jaffel 2010; Ehrenreich et al. 2008*). The absorption in Lyα was also observed at the hot Jupiter HD189733b (*Lecavelier des Etangs et al. 2010; Bourrier and Lecavelier des Etangs 2013*) and a colder giant 55 Cnc b (*Ehrenreich et al. 2012*). The measurements at transit of warm Neptune GJ 436b (*Ehrenreich et al. 2015*) revealed 50%, mostly blue-shifted, absorption well beyond the measurement uncertainty. Altogether these observations indicate the existence of an essentially dynamic and expanding hot hydrogen envelope around the close-orbit exoplanets, driven by complex processes of stellar planetary interaction and radiative energy deposition. The presence of an outflowing exosphere is also qualitatively supported by observation of heavier element lines, such as carbon, oxygen, and silicon at HD 209458b (*Vidal-Madjar et al., 2004; Linsky et al., 2010*), which show the absorption beyond the Roche lobe at velocities up to 50 km/s in blue and red wings of the corresponding lines (*Ben-Jaffel 2010*), as well as the measurements of Si III absorption during a transit of HD 189733 b (*Bourrier et al., 2013*) and MgII signatures in WASP-12 b (*Fossati et al., 2010*) found as significant near-UV absorption.

Soon after discovery of first hot Jupiters, the 1D gasdynamic models have been developed to describe their exospheres (*Yelle 2004, Tian et al. 2005, García Muñoz 2007, Penz et al. 2008, Murray-Clay et al. 2009, Koskinen et al. 2010, Guo 2011, Trammell et al. 2011, Shaikhislamov at al. 2014*). It has been found, that due to extreme heating by the ionizing XUV stellar radiation, the close-orbit planets should indeed possess an atmospheric material (partially ionized gas) outflow in the form of a super-sonic planetary wind (further PW) which overcomes the planet's gravity and expands beyond the Roche lobe. The calculated exosphere temperature of $\sim 10^4$ K, outflow velocity of ~10 km/s, and mass loss in the range of $\sim 10^9$–$10^{11}$ g/s qualitatively agree with transit spectral observations. The major factors and processes influencing the formation of PW, such as the spectra of ionizing stellar XUV radiation and particulars of its absorption, gravitational/tidal effects, hydrogen plasma- and photo- chemistry, effects of heavier species in atmospheric composition, processes of infrared and ultraviolet cooling, have been analyzed in these works.

With the detection of absorption by HD209458b at high velocity wings of stellar Lyα it has been proposed that ENAs are produced due to the acceleration of hydrogen atoms by the radiation pressure of Lyα in the thermosphere of a planet, expanded beyond the Roche lobe (*Vidal-Madjar et al. 2003, Lecavelier des Etangs et al. 2004, Lecavelier des Etangs et al. 2008*). It appears that the radiation pressure acting on a hydrogen atom is several times higher than the stellar gravity. Thus, it is able to accelerate the atom significantly during the time-span of its ionization. At the same time, as it was demonstrated by *Erkaev et al. (2005), Khodachenko et al. (2007a,b), and Lammer et al. (2009)*, hot exosphere of exoplanets should experience a non-thermal escape of particles. As proposed by *Holmström et al. (2008)* and *Ekenbäck et al. (2010)*, the escaping planetary atoms are ionized in the flow of stellar wind (further SW) protons via charge-exchange reaction. This mechanism produces fast hydrogen atoms (i.e., ENAs), necessary to explain the asymmetric absorption in the blue spectral wing of Lyα, not by acceleration of particles, but due to the presence in the SW of fast and hot protons. Nowadays, for the simulation of such ENA generation process, the kinetic models of test particles are widely applied, that take into account both the effects of radiation pressure, and charge-exchange. This approach has been realized, in particular, for the interpretation of transit depletion in Lyα for HD209458b (*Kislyakova et al. 2014*), HD189733b (*Bourrier and Lecavelier des Etangs 2013*), and GJ 436b (*Bourrier et al. 2015*). However, significant limitation of such modeling consists in the launching of planetary particles into SW from a semi-empirically pre-defined boundary of the planetary exosphere, without taking into account the realities of the PW formation, structure, and plasma flows (i.e., PW and SW) interaction.

In the meantime, a concept of HD/MHD modelling of the escaping PW and its interaction with the SW was developing. *Murray-Clay et al. (2009)* estimated that for HD209458b, the typical SW ram pressure, based on the solar wind parameters, equals the PW thermal and ram pressures at a distance of about 4-5 planetary radii, $R_p$, which is close to the point, where the PW flow becomes a super-sonic one. This conclusion has been confirmed later in our simulations (*Shaikhislamov et al. 2014, Khodachenko et al. 2015*), which provided the location of the pressure balance point beyond the

Roche lobe at the distance ~5 $R_p$. Despite of the obvious necessity to merge in a proper way the above mentioned HD/MHD and kinetic test-particle models in order to build the complete and consistent view of the PW initiation, propulsion and consequent interaction with the SW, such attempts still remain quite rare and of limited character. One of the reasons for that is a complication of the modelling approach, which requires switching from the 1D to at least 2D geometry, separate description of the interacting and essentially different plasma components (i.e. PW and SW partially ionized plasmas), appropriate account and combining of the large-scale processes of plasma flows interaction and micro-processes of the PW initiation and propulsion at atmospheric scale height. There is a number of works, where some of the above mentioned complexity aspects have been addressed by adapting of various astrophysical codes. In particular, a shock wave at PW and SW collision region has been modeled in *Stone and Proga (2009)*. Further on, a 3D picture of planetary material, overflowing the Roche lobe and falling onto the star by irregular clamps, has been simulated in *Bisikalo et al (2013)* for the very close exoplanets like Wasp-12b. In course of the parametric study at different SW conditions performed by *Matsakos et al.* (*2015*), it has been shown that, if the SW stops the planetary material flow inside the Roche lobe, then the material flow structure with a comet-like trailing tail is formed. In the opposite case, an escaping PW stream is formed at the day- and night- sides, with the dayside stream accretion on the star, accompanied by its fast loss of the rotational momentum due to interaction with SW. It has been proposed in *Tremblin & Chiang* (*2013*) for the first time to consider within the frame of HD description the formation of an ENA cloud around a hydrogen rich exoplanet by modeling of four interconnected fluids: 1) planetary hydrogen atoms, 2) protons of planetary origin, 3) stellar protons, and 4) ENAs. However, in this work, as well as in *Christie et al. 2016*, besides of numerous empirical simplifications, such approach was not realized in full measure, because the motion of planetary hydrogen atom and proton fluids was not distinguished. This lead to erroneous conclusions, resulting that charge-exchange and production of ENAs was possible only in a thin boundary layer of turbulent mixing. In the last years some of the previously used 1D fluid models have been upgraded to 3D ones, which still retained ungrounded simplifications of the physics of the PW formation (for example, *Tripathi et al. 2015*).

In our previous paper *Shaikhislamov et al.* (*2016*) we performed a 2D four fluid simulation of PW and SW interaction for an unmagnetized analog of the tidally locked exoplanet HD209458b. Because of the completely self-consistent approach developed in *Shaikhislamov et al.* (*2014*) and *Khodachenko et al.* (*2015*), which doesn't rely on any quasi-empirical simplifying assumptions, while employing an independent description of planetary hydrogen atoms and protons, interacting with stellar protons and ENAs, we were able to combine the micro-physics of PW generation with its large-scale expansion and interaction with SW. This enabled us making (on the base of obtained results) of several important conclusions. First, depending on natural (about an order of magnitude) variations in the dynamic pressure of the solar type SW, the PW of HD209458b can exist in two essentially different regimes of its outflow (*Shaikhislamov et al. 2016*): 1) the *"Blown by the wind"* regime, when sufficiently strong SW confines the escaping PW at the day-side and channels it away from the star into the tail, forming a kind of an elongated planetary plasmasphere, and 2) the *"captured by the star"* regime, when the tidal force exceeds the action of the SW ram pressure and a stream-type structure of the escaping PW is formed along the planet-star line (in the tail-ward and towards the star). The planetary material flow is sufficiently dense to remain strongly collisional even rather far from the planet (several tens of $R_p$). In both regimes, a thin boundary, or a kind of ionopause, between the PW and SW plasmas is formed, at which the corresponding PW and SW plasma pressures are balanced. However, while the planetary protons are fully stopped or redirected, at the ionopause, atoms can penetrate through it into the SW where they undergo charge-exchange reaction and generate ENAs. By this, in the case of the *"captured by the star"* regime the extended boundary between the quasi-parallel PW stream and the SW flow is distorted by a kind of interchange instability which generates a turbulent vortex layer. It appears that more ENAs are generated under the conditions of the *"blown by the wind"* regime of the PW and SW interaction, because in this case a shocked region is formed beyond the ionopause (*Shaikhislamov et al. 2016*). However, we found that for the used parameters of the model, such as the Solar type XUV flux of 4.466 erg cm$^{-2}$ s$^{-1}$ at 1 a.u. and the typical for the Sun density, velocity and temperature of the SW resulting in a pressure of about $5 \cdot 10^{-6}$ µbar the amount of ENAs, obtained in simulations of HD209458b, appears an order of magnitude smaller than needed to influence the transit observations in Lyα line.

In the present paper we analyze in details the absorption of Lyα line in the dynamical environment of the interacting PW and SW around HD209458b with the purpose to understand what information can be derived from the planet's transit measurements in EUV. In particular, we look for the conditions, at which the absorption is more or less symmetric over blue and red wings of the Lyα line caused by the *non-resonant natural line broadening* within the dense exosphere of the planet inside the Roche lobe, as well as for the conditions at which the absorption is asymmetric and caused by the *resonant*, or *thermal line broadening* produced by the ENAs, increasing the depth of the transit in the blue shifted wing. To make our model relevant to the case under study, we first of all, included in it the Lyα radiation pressure as one of the factors influencing the generation and distribution of ENAs. Then, we varied by more than an order of magnitude such parameters of the model as the stellar XUV flux and SW ram pressure (controlled by SW density and velocity). It has been found that under the explored conditions and Lyα flux based on actual measurements, the radiation pressure cannot be a dominating factor that explains the observations. At the PW densities predicted by the simulations its influence on the formation of ENA cloud around HD209458b is small in comparison with charge-exchange. We support these conclusions by physical explanations and quantitative estimations.

The paper is organized in the following way. In Section 2 we describe the used model for the simulation of PW and SW interaction, paying attention to the basic processes, equations, and important details regarding the formation of Lyα absorption line. In Section 3 the results of numerical simulations of the HD209458b case under different XUV and SW conditions, and the related peculiarities of the Lyα absorption line expected during the planet transits, are presented. In particular Sections 3.1 and 3.2 are dedicated to the study of the role of the *resonant thermal* and *non-resonant natural* line broadening absorption, the contribution of ENAs, and the influence of XUV flux on the whole Lyα absorption profile. A comparative study of the radiation pressure and charge-exchange, as two major mechanisms for the production of ENAs in the vicinity of the planet is presented in Section 3.3, whereas the Lyα absorption features expected in the case of the "*blown by the wind*" regime of PW and SW interaction are investigated in Section 3.4. Section 4 is dedicated to the discussion of the obtained results and conclusions.

## 2. Model

Our model has been described in previous works by *Shaikhislamov et al.* (*2016*) and *Khodachenko et al.* (*2015*), where the details regarding physical processes, equations, hydrogen photo-chemistry, problem geometry, numerical code implementation, boundary conditions, and imposed approximations, as well as their justification, are provided. In that respect, here we just address (and repeat) the most essential points. To simulate the interaction of PW and SW, we apply a multi-fluid model which includes, as a separate fluids, the hydrogen and helium components of the planetary origin ($H_2, H_2^+, H_3^+, H, H^+, He, He^+$), as well as the hydrogen atoms (ENAs) and protons of the SW origin ($H_{SW}, H_{SW}^+$). Besides of charge-exchange all particles exchange momentum and temperature through the elastic collisions. The modeled system of HD209458b consists of a planet with mass $M_p = 0.71 M_J$ and radius of $R_p = 1.38 R_J$ orbiting at the distance of D=0.047 a.u. around a solar-type G-star with the mass of $M_{st} = 1.148 M_{Sun}$ and radius of $R_{st} = 1.13 R_{Sun}$. The planetary and stellar radius and mass are scaled in the units of radius and mass of the solar system Jupiter ($R_J =$

71.5 × 10³ km, $M_J$ = 1.89 × 10²⁷ kg) and the Sun ($R_{Sun}$ = 69.6 × 10⁴ km, $M_{Sun}$ = 1.99 × 10³⁰ kg), respectively. As an initial state, the neutral atmosphere in a barometric equilibrium with the base temperature of 1000 K is taken, that consists of molecular hydrogen and Helium at partial ratio $x_{He}/x_{H2} = 1/5$. The radiative energy deposition is calculated by the spectral integration of XUV flux, for which we use the solar proxy spectrum in the range 10–912 Å, compiled by *Tobiska (1993)* and binned by 1 Å. The total integrated XUV flux at 1 a.u. for this spectrum is $F_{XUV}$=4.466 erg s⁻¹ cm⁻². For the investigation of the effect of varying XUV flux on the absorption at high velocity blue wing of the stellar Lyα line during HD209458b transit, the proportionally reduced/increased values of the basic solar proxy flux were used. As reference parameters of typical SW at 0.047 a.u., we use the simulated data from *Johnstone et al. (2015)* and consider the *slow* and *fast* SW with the following ranges of density $n_{sw}$ = 4600...2.5×10⁴ cm⁻³, velocity $V_{sw}$ = 230...500 km/s, and temperature $T_{sw}$ =(1.2...2.9) ×10⁶ K, which correspond the total SW pressure range of $p_{sw} = 5 \cdot 10^{-6}...1.3 \cdot 10^{-4} \mu bar$. To employ our 2D axially-symmetric hydrodynamic (HD) numerical model with the symmetry axis taken along the planet-star line and with the center of reference attached to the center of the planet (*Shaikhislamov et al. 2016*, *Khodachenko et al. 2015*), we disregard the Coriolis force and ignore the transverse component of the SW velocity, related with the orbital motion of the planet around the star, which is ~140 km/s for the HD209458b. As shown in *Shaikhislamov et al. (2016)*, the last assumption is still possible for the considered planet in both cases; of *fast* and *slow* SW velocities. As the characteristic values of numerical problem we use the planet radius $R_p$, a temperature of 10⁴ K and corresponding thermal velocity of ions $V_o = 9.07 \, km/s$.

The interpenetration of counter-streaming plasma flows of PW and SW is restricted by gyro-rotation in the background magnetic field. The frozen-in magnetic fields, either laminar or chaotic, are always present in both stellar and planetary plasmas. As compared to the characteristic scale of the considered system $\sim R_p$ ($>10^{10}$cm), even for a very weak magnetic field of $1\, nT$, the proton gyroradius is at least ten times smaller than the characteristic scale. Thus, under the natural conditions of the colliding PW and SW by HD209458b the interpenetration of planetary and stellar protons should be microscopically small. Because of that, the interaction between the planetary and stellar protons was described in *Shaikhislamov et al. (2016)* by effective collisions with a very small mean free path (about a particle gyroradius). Such a strong coupling means that proton fluids of planetary and stellar origin have the same velocity and temperature whenever they mix. In view of that, in the present paper we simplify the model by using a physically equivalent *single proton fluid* which is a subject to two different boundary conditions – at the planet's surface (zero density and velocity) and at the outer boundary (SW parameters).

An important modification of the model consists in inclusion of the radiation pressure by Lyα photons. The reconstructed data on the Lyα line profile based on actual measurements for the HD209458 can be found in *Wood et al. (2005)* or in *Burrier and Lecavelier des Etangs (2013)*. In our simulations, we use its simplified analytical version, in which it has a constant value of $F_{Ly\alpha,\lambda} = 8 \cdot 10^3 \, erg \, cm^{-2} s^{-1} Å^{-1}$ in the range of Doppler shifted velocities $\pm 45 \, km/s$, and drops outside this range linearly to zero at the velocities $\pm 140 \, km/s$. This flux (with a total value $\approx 13 \, erg \, cm^2 s^{-1}$ at 1 a.u.) is absorbed with the specified below integrated cross-section of $\sigma_{Ly\alpha,\lambda} = \int \sigma_{abs} d\lambda \approx 5.5 \cdot 10^{-15} cm^2 \cdot Å$. That generates at the center of the emission line a force $f_{rad} = F_{Ly\alpha,\lambda} \cdot \sigma_{Ly\alpha,\lambda}/c$ which is about 3.5 times larger than the stellar gravity pull and consistent with the value calculated, e.g., in *Burrier and Lecavelier des Etangs (2013)*.

The spectral absorption of Lyα photons is described assuming the so called Voigt convolution of a Lorentz line shape with a natural width and Gaussian distribution profile, which depends on temperature T, velocity along the line of sight $V_z$, and hydrogen atom density $n_H$:

$$\text{Absorption} = 1 - \frac{I_{\text{transit},v}}{I_{\text{out},v}} = \frac{2}{R_{St}^2} \int_0^{R_{St}} (1 - e^{-\tau}) \cdot r\, dr, \qquad (1)$$

where $\tau(V) = \int_{-L}^{L} dz \cdot n_H \cdot \sigma_{abs}(V, V_z, T)$ is an optical depth and $\sigma_{abs}(\text{cm}^2) = f_{12}\sqrt{\pi}\frac{e^2}{m_e c^2} \cdot \frac{c}{\Delta v_D} \cdot H = f_{12}\sqrt{\pi}\frac{e^2 \lambda_o}{m_e c^2}\sqrt{\frac{m_p c^2}{2kT}} \cdot H = 5.9 \cdot 10^{-14} \cdot \sqrt{10^4\,\text{K}/T} \cdot H$ is absorption cross-section, where $H = \frac{1}{\pi} \cdot \alpha \int \frac{e^{-y^2}}{(x-y)^2 + \alpha^2} dy$, is the Voigt profile, which takes into account the convolution of two broadening mechanisms, one of which alone would produce a Gaussian profile (as a result of the Doppler broadening), and the other would produce a Lorentzian profile. Here $\alpha = \Delta v_L / 2\Delta v_D = \frac{\Delta v_L}{v_o}\sqrt{\frac{m_p c^2}{8kT}}$ and $x = \frac{V - V_z}{\sqrt{2kT/m_p}}$, $f_{12}$ is the oscillator strength, $\Delta v_L$ and $\Delta v_D$ are the Lornz and Dopler line widths, respectively.

Instead of wavelength, the absorption is expressed in terms of Doppler shifted velocity in the reference frame of the star. The averaging in equation (1) covers the stellar disk of radius $R_{St}$. The mean fluid values of $n_H, T, V_z$ are calculated by the numerical model. As the length L we take an empirical value of L=10$R_p$. This length is limited by the spiraling of planetary material stream due to Coriolis force (clockwise in the planet based reference frame) which is not included in our model. In other words, L corresponds approximately to the size of our modelling box, where the spiraling might still be neglected. It can be considered also as the width of the PW streams, appeared to be roughly of ~10$R_p$. The assumed size of the neglecting by spiraling is also consistent with the large-scale 3D simulations by *Kislyakova et al.* (*2014*) and *Burrier et al.* (*2013*).

At temperatures above 2 K, which is a definitely satisfied condition in our case, the Voigt profile integral can be fitted (with accuracy better than 1%) by an analytical expression (*Tasitsiomi 2013*):

$$H \approx \exp(-x^2) + \frac{\alpha}{x^2 \sqrt{\pi}} \cdot q(x^2),$$

where

$q(x^2) = \frac{21 + x^2}{1 + x^2} \cdot z(x^2) \cdot [0.1117 + z(x^2) \cdot [4.421 + z(x^2) \cdot (5.674 \cdot z(x^2) - 9.207)]], \quad z(x^2) > 0$

$q(x^2) = 0, \quad z(x^2) < 0$

and

$z(x^2) = (x^2 - 0.855)/(x^2 + 3.42)$, so that $q(\infty) = 1$, $q_{max} = q(3.865) = 1.62$

Therefore, the corresponding absorption cross-section can be approximated as follows:

$$\sigma_{abs}(\text{cm}^2) \approx 5.9 \cdot 10^{-14} \cdot \sqrt{10^4\,\text{K}/T} \cdot \exp(-x^2) + 2.6 \cdot 10^{-19} \cdot \left(\frac{100\,\text{km/s}}{V - V_z}\right)^2 \cdot q(x^2) \qquad (2)$$

The applied analytical fit explicitly shows that there are two different kinds of the Lyα absorption, caused by essentially different populations of particles. First, is the *resonant*, or *thermal line broadening* absorption related with resonant atoms matching the Doppler shifted velocity of the line profile which, at the typical temperature of order of $10^4$ K, has a significant cross-section of the order of $10^{-13} cm^2$ (first term in equation (2) for the absorption cross-section). And second, is the *non-resonant natural line broadening* absorption related with far wings of Lorentz line profile with a much smaller cross-section, which at velocity of 100 km/s is of the order of $10^{-19} cm^2$ (second term in equation (2)). Note that the same formula, albeit differently normalized and without correcting factor $q(x^2)$, is widely used in literature. In our simulations we apply the whole equation (2) to calculate Lyα line absorption in order to account for self-shielding of radiation pressure.

## 3. Results

### 3.1 PW and SW interaction under typical conditions of HD209458b, and the related Lyα absorption features.

The modelled structure of escaping PW of HD209458b interacting with the SW flow is shown in color plots of Figure 1. The total integrated XUV flux $F_{XUV}$ in this simulation run was taken to be 4.466 erg cm$^{-2}$ s$^{-1}$ at 1 a.u., while the SW parameters were $n_{SW} = 4600 cm^{-3}$, $V_{SW} = 230 km/s$, $T_{SW} = 1.27 MK$, resulting in the total SW pressure $p_{SW} = 5 \cdot 10^{-6} \mu bar$. These values are typical for the so-called *slow* wind of a solar type star. As it has been shown in our previous paper *Shaikhislamov et al. (2016)*, the *"captured by the star"* regime of PW and SW interaction is realized for such parameters, when, besides of a tail-ward planetary material outflow, the partially ionized plasma stream pulled by the stellar gravity is formed on the dayside of the planet. This dayside stream is surrounded by the counter-streaming SW plasma, and the whole planetary plasmasphere is separated from the SW by a thin ionopause layer, at which sporadic interchange instabilities develop. While the expansion of the PW protons across the stream is stopped by the thermal pressure of the SW protons, the planetary atoms penetrate through the ionopause and are injected into the SW where they are accelerated by the pressure gradient and radiation pressure, while undergoing the charge-exchange. These processes result in the generation of ENAs. The ENAs are mostly produced in vicinity of the ionopause where the density of atoms is the largest. Both ENAs and planetary atoms expand further across the stream and are gradually swept away from the star. As it can be seen in the streamlines plot in Figure 1, only a small portion of the total amount of the escaped planetary atoms passes sufficiently close to the ionopause to be able to penetrate into the SW (mainly due to the boundary instabilities and pressure gradient).

The absorption profile of Lyα line revealed in this simulation is given in Figure 2. In particular, the absorption is rather significant at low Doppler-shifted velocities, i.e. within the range of ±50 km/s. However, this range of velocities cannot be probed with the existing observations, due to the geo-coronal contamination and interstellar extinction of the line. At the same time, one can see that at blue and red shifted velocities beyond ±100 km/s the absorption reaches values of about 6%, which are in the range of those actually detected. An important insight into the nature of the absorption line profile can be derived from the analysis of the different kinds of Lyα absorption related with different contribution of different particle populations in the modelled system. In the Figure 2 the calculated absorption due to resonant atoms (the part of *resonant thermal line broadening*) is separately shown. By subtracting it from the total absorption the *non-resonant natural line broadening* part can be found in the first approximation. It is also shown in Figure 2. In a similar way, the decomposition to distinguish between the contributions of ENAs and planetary atoms is made. Note that the absorption by the planet disk, equal to 0.0156%, is not included neither in the resonant, nor in ENA parts. One can see in Figure 2 that the strong absorption within the ±50 km/s interval of velocities is produced by the resonant planetary atoms in the PW streams moving towards and outwards the star (along the observation line). According to the simulation, at distances of ~10$R_p$ from the planet the escaping PW streams are accelerated up to the velocities of about 50 km/s (for details see also in *Shaikhislamov et al. 2016*) resulting in the *resonant thermal line broadening*. At larger Doppler-shifted velocities, the absorption by the planetary atoms due to *non-resonant natural line broadening* becomes dominant. It has been found that this absorption is produced by an extended, but still sufficiently dense, planetary exosphere with $n_H > 10^7 \text{cm}^{-3}$ at relatively close distances from the planet within the Roche lobe. For example, 90% of it comes from the region approximately bounded by a cylinder with $r = 2.5 R_p$, elongated along the planet-star line over $-3R_p < Z < 3R_p$. Due to relatively low bulk velocity of material in this region, the absorption due to *non-resonant natural line broadening* is symmetric and gives practically the same contribution at both the red and the blue wings of the Lyα line.

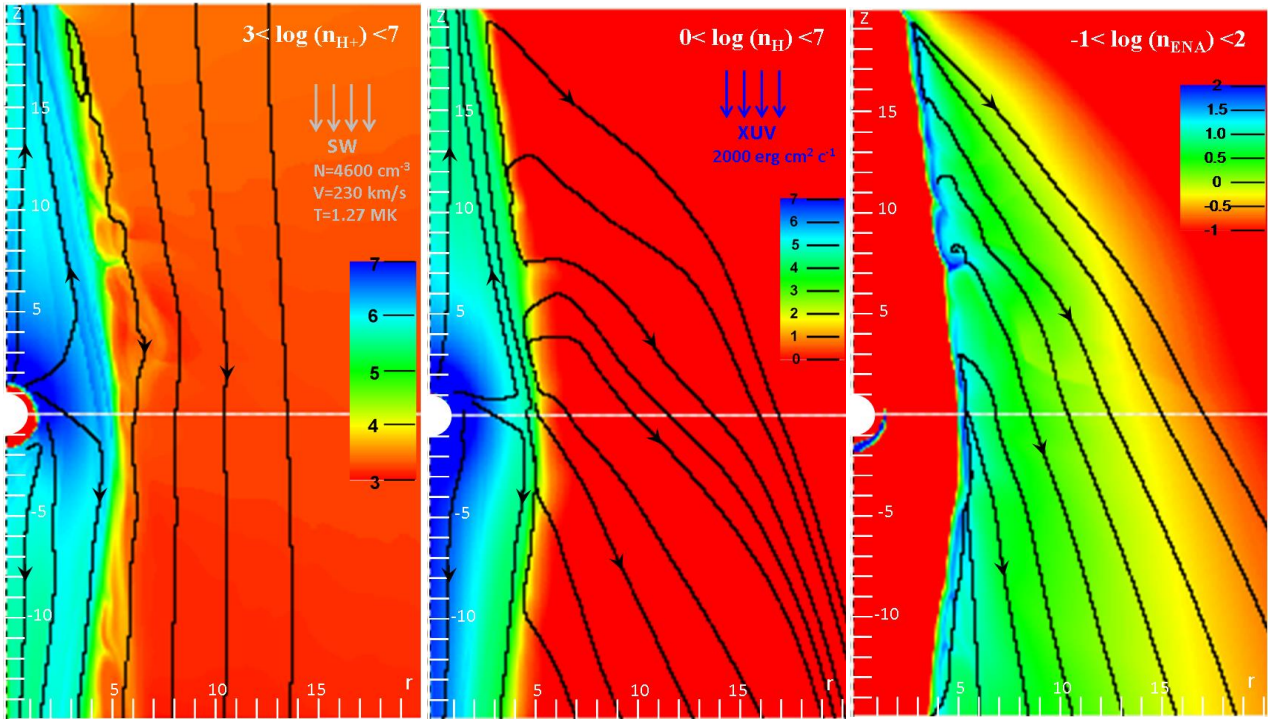

**Figure 1.** Density distributions of protons ($n_{H+}$), planetary atoms ($n_H$) and ENAs ($n_{ENA}$), realized in the *"captured by the star"* regime of PW and SW interaction around HD209458b for the total XUV flux at 1 a.u. $F_{XUV}$=4.466 erg s$^{-1}$ cm$^{-2}$ and under the conditions of *slow* SW with the total pressure $p_{sw} = 5 \cdot 10^{-6} \mu\text{bar}$. Here, and further on in similar plots, white circle indicates the planet; the plotted values are in log scale; the streamlines of the corresponding components are shown in black; the values outside the indicated variation ranges of the plotted parameters are colored either in red if smaller than minimum, or in blue, if higher than maximum.

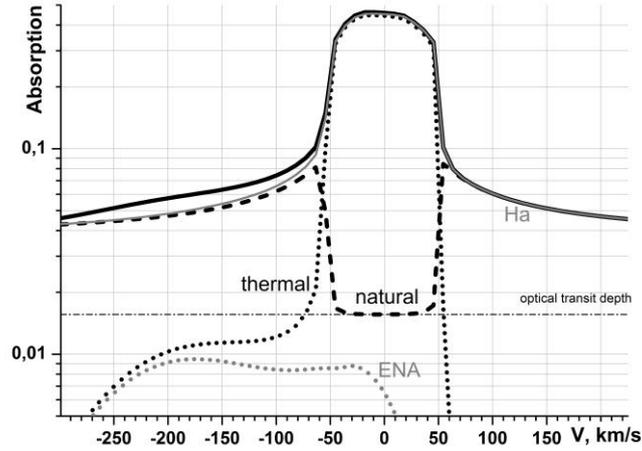

**Figure 2.** The absorption profile of Lyα line (thick solid line) which corresponds to the simulations presented in Figure 1 (*slow* SW; *"captured by the star"* regime of interaction). A decomposition of the total absorption onto the *resonant thermal* (black dotted) and *non-resonant natural line broadening* (black dashed) parts is shown. Another decomposition represents the contributions to the absorption of ENAs (gray dotted) and planetary atomic hydrogen Ha (gray solid).

This conclusion is in full agreement with the analysis by *Ben-Jaffel* (*2007, 2008*), who found that the absorption measured during the transits of HD209458b is about the same in the blue and red parts of Lyα line. It is also in agreement with its physical interpretation given in *Ben-Jaffel* (*2007, 2008*), as well as in *Koskinen et al. (2010)*, according to which the absorption is caused by the atoms of an expanded thermosphere confined inside the Roche lobe. In particular, from the set of three transits, measured with resolution of 0.08 Å (*Vidal-Madjar et al. 2003*), the inferred full-width transit depth in statistically reliable blue and red parts of the Lyα line ($-207 \text{km/s} < V < -77 \text{km/s}$ and $54 \text{km/s} < V < 188 \text{km/s}$ in Doppler shifted velocity units is about $(8.4 \div 8.9) \pm 2\%$ (*Ben-Jaffel 2007, 2008*), while the set of four transits measured with resolution of 2.5 Å (*Vidal-Madjar et al. 2004*) gave a lower value of $6.6 \pm 2.2\%$ (*Ben-Jaffel 2010*). Using a 1D numerical model, *Koskinen et al.* (*2010*) estimated the transit depth at the level of 6.6%.

In our simulation, after averaging over the same line intervals as reported in *Ben-Jaffel* (*2007*) we obtain a slightly lower transit depth value of 6%. In our simulation, the transit depth calculated at the same line intervals as those used in *Ben-Jaffel* (*2007*) for medium resolution mode, gives 6.6% and 6.3% at $F_{XUV}=4$ erg s$^{-1}$ cm$^{-2}$ and $F_{XUV}=8$ erg s$^{-1}$ cm$^{-2}$, respectively. The absorption over the whole line (excluding the core $-64 \text{km/s} < V < 42 \text{km/s}$) gives about the same values, i.e. $(6.1 \div 6.3)\%$.

An interesting feature is that the absorption due to *non-resonant natural line broadening* doesn't depend on XUV intensity and the total mass loss in a wide range of values from $\dot{M} \sim 4 \cdot 10^{10}$ g/s at $F_{XUV} \sim 1 \text{erg cm}^{-2}\text{s}^{-1}$ upto $\dot{M} \sim 4 \cdot 10^{11}$ g/s at $F_{XUV} \sim 10 \text{erg cm}^{-2}\text{s}^{-1}$. This is because the temperature of the escaping material in close vicinity of the planet changes with XUV rather weakly, while the flow velocity, which defines the mass loss, builds up at several planetary radii. Thus, the densest exosphere at distances below $(1 \div 2)R_p$ around the planet which contributes the Lyα *non-resonant natural line broadening* absorption remains mostly in a barometric equilibrium and is not affected by the PW outflow at higher altitudes. However, we found that the hydrogen chemistry somewhat affects the transit depth caused by the *non-resonant natural line broadening* absorption. For example, taking purely atomic hydrogen atmosphere (instead of molecular one) decreases absorption by about 0.5%. This is because of dissociation which makes the atomic hydrogen more abundant in the molecular atmosphere up to the H2/H dissociation front, which in our simulations is located at about 0.3 μbar.

**3.2 On the Lyα *resonant thermal line broadening absorption* controlled by stellar XUV**

Regarding the *resonant thermal line broadening* absorption of Lyα, one can see in Figure 2 that at high velocities, i.e. for $|V|>60$ km/s it is an order of magnitude lower than the *non-resonant natural line broadening* absorption. Therefore, as has been found in our previous paper *Shaikhislamov et al. (2016)* and confirmed in the present work, for the typical for a Sun like star XUV flux and SW conditions at the distance of D=0.047 a.u., the amount of ENAs generated due to charge-exchange and the radiation pressure is too small to produce any detectable asymmetry between the red and blue wings of the absorption line during the transits of HD209458b. To explain such small population of ENAs, the following important physics points were emphasized in *Shaikhislamov et al. (2016)*.

First, the escaping PW either as a supersonic stream or an expanding cloud, surrounding the planet, is strongly collisional due to proton-atom resonant charge-exchange. While the colliding with each other planetary and stellar proton flows form a thin contact ionopause, at which their respective pressures balance each other, the planetary atoms on their way towards the region filled with the SW plasma have to drag through the volume occupied by planetary protons. Since the mean-free path of momentum exchange in this case appears to be significantly smaller than the typical size of plasmosphere, the relative atom-proton velocity is proportionally reduced in comparison with that of the thermal motion. In particular, assuming an approximate pressure - friction force balance: $\frac{\Delta p_H}{\ell} \sim m_p n_H n_{H+} \sigma_{exch} v_T \Delta V_{H,H+}$, where $\Delta p_H \sim n_H T$, the relative atomic flux passing through the ionopause into the SW region, can be estimated as follows:

$$n_H \Delta V_{H,H+} \sim n_H v_T / (n_{H+} \sigma_{exch} \ell) \qquad (2)$$

Thus, taking the typical size as the radius of the escaping PW stream $\ell \sim 5 R_p$ and proton density $n_{H+} \sim 3 \cdot 10^5 \text{cm}^{-3}$, we obtain at $\sigma_{exch} \approx 5 \cdot 10^{-15} \text{cm}^2$ that $\Delta V_{H,H+} \sim 10^{-2} v_T$.

Second, only those planetary hydrogen atoms which are able to reach the SW, will generate the ENAs. From equation (2) one can see also that the flux of injected into the SW region atoms is proportional to the ratio of densities $n_H \Delta V_{H,H+} \sim n_H / n_{H+}$. This ratio is determined mostly by the ionization of atoms due to the stellar XUV and in a less degree – by recombination. In the first approximation, the number of atoms can be supposed decreasing exponentially with time, as they stream away from the planet, and the typical ionization time is proportional to the total XUV flux. For the typical XUV flux estimated for a Sun like star at the orbital distance of HD209458b, this results in the indeed small amount of planetary hydrogen atoms reaching the SW region. It may be, however, expected that the decrease of XUV intensity would result in a rapid increase of the number of atoms that survive the exposure to the ionizing radiation before they penetrate into the SW. The decrease of XUV flux also results in total decrease of temperature of the escaping PW and, consequently, in a faster recombination, again increasing the ratio $n_H / n_{H+}$.

We note that the above mentioned circumstances are important also for the acceleration of hydrogen atoms by the radiation pressure, and have to be properly taken into account. In particular, the PW flow is sufficiently dense to be shielded from Lyα photons and the momentum received by atoms is additionally reduced because it is shared with much more numerous protons (*Shaikhislamov et al. 2016*). As it is empirically known for the Sun, its XUV flux is highly variable in time. Even the annually averaged value changes by about three times between the solar minimum and maximum phases. Therefore, a possibility of similar variations of XUV flux on a Sun like star HD209458 should be taken into consideration. In order to check, whether the reducing of the XUV flux might increase the absorption at high velocity blue wing of the stellar Lyα line during HD209458b transit, we performed a set of dedicated simulations. Figure 3 shows the transit profiles of the total Lyα absorption (left panel) and its *resonant thermal line broadening* part (right panel) for different values of XUV flux, decreasing by twofold from 16 erg cm$^{-2}$ s$^{-1}$ to 1 erg cm$^{-2}$ s$^{-1}$. According to the simulation results, the amount of ENAs steeply increases with the decrease of XUV flux, and one can see in Figure 3, that indeed, the Lyα absorption in blue-shifted velocity wing reaches the level of ~15% at the XUV flux value of 1 erg cm$^{-2}$ s$^{-1}$. At the same time, the absorption in red wing at velocities above 60 km/s, which is produced by *non-resonant natural line broadening*, remains practically independent on XUV flux. It is worth to mention, that at XUV flux of 4 erg cm$^{-2}$ s$^{-1}$ and higher, the absorption profile is very much symmetric, while at lower fluxes a strong asymmetry of the absorption line develops with a significantly increased absorption in blue wing.

To illustrate in detail the influence of XUV on the generation of ENAs, we show in Figure 4 the profiles across the PW stream of hydrogen atom to proton density ratio $n_H/n_{H+}$ and transverse atom-proton relative flux $n_H \Delta V_{H,H+}$ (see Equation (2)) at distance Z=5R$_p$ upstream from the planet for different values of XUV flux. These plots demonstrate that the ratio $n_H/n_{H+}$ and hydrogen atoms flux injected into the SW area vary roughly in an inverse proportion with XUV intensity. Figure 4 shows also that there is a region adjacent to the planetary stream boundary where atoms have a distinct radial velocity relative to protons which varies from $\sim 10^{-3} V_o$ for the XUV4 case up to $\sim 10^{-2} V_o$ for the XUV1 case. The PW stream width predictably contracts from about 5R$_p$ to about 3R$_p$ for XUV4 and XUV1 cases, respectively, because it is controlled by the thermal pressure which depends on the XUV heating.

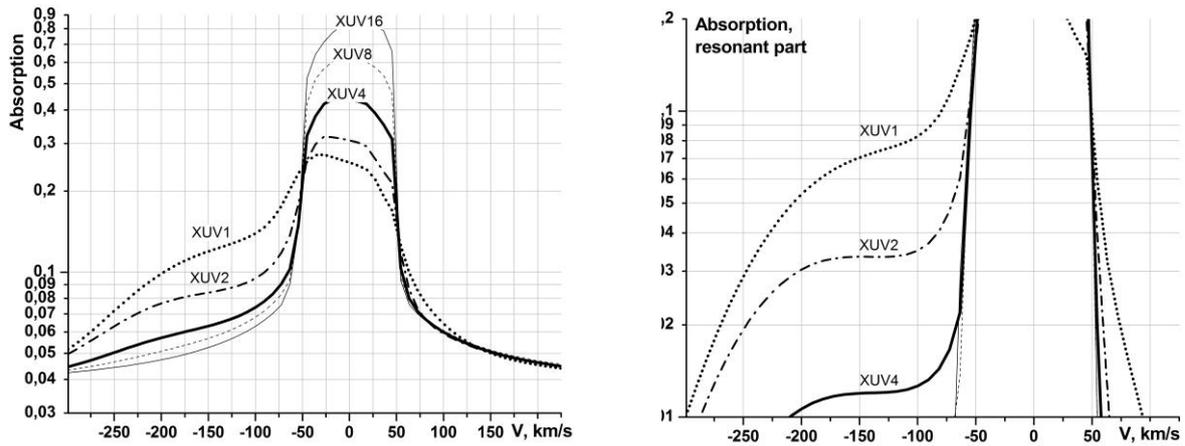

**Figure 3.** Absorption profiles of Lyα line for different values of XUV flux F$_{XUV}$, given at 1 a.u.: 16 erg cm$^{-2}$ s$^{-1}$ (thin solid), 8 erg cm$^{-2}$ s$^{-1}$ (dashed), 4 erg cm$^{-2}$ s$^{-1}$ (thick solid), 2 erg cm$^{-2}$ s$^{-1}$ (dash-dot), 1 erg cm$^{-2}$ s$^{-1}$ (dotted). <u>Left panel</u>: the total absorption profile; <u>Right panel</u>: the absorption part due to the *resonant thermal line broadening*.

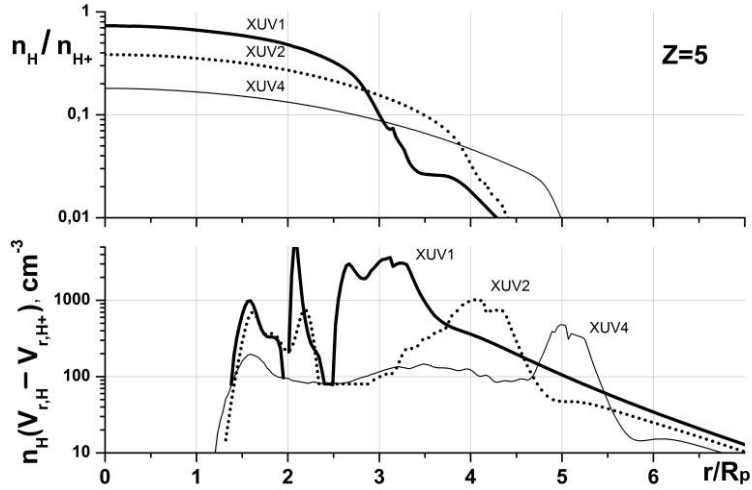

**Figure 4.** Cross-stream profiles of hydrogen atom to proton density ratio (upper graph) and transverse atom-proton relative flux (bottom graph) at upstream distance $Z=5R_p$ from the planet for different values of XUV flux, given at 1 a.u.: 4 erg cm$^{-2}$ s$^{-1}$ (thin solid), 2 erg cm$^{-2}$ s$^{-1}$ (dotted), 1 erg cm$^{-2}$ s$^{-1}$ (thick solid). Velocity is taken in units of $V_o = 9.07$ km/s.

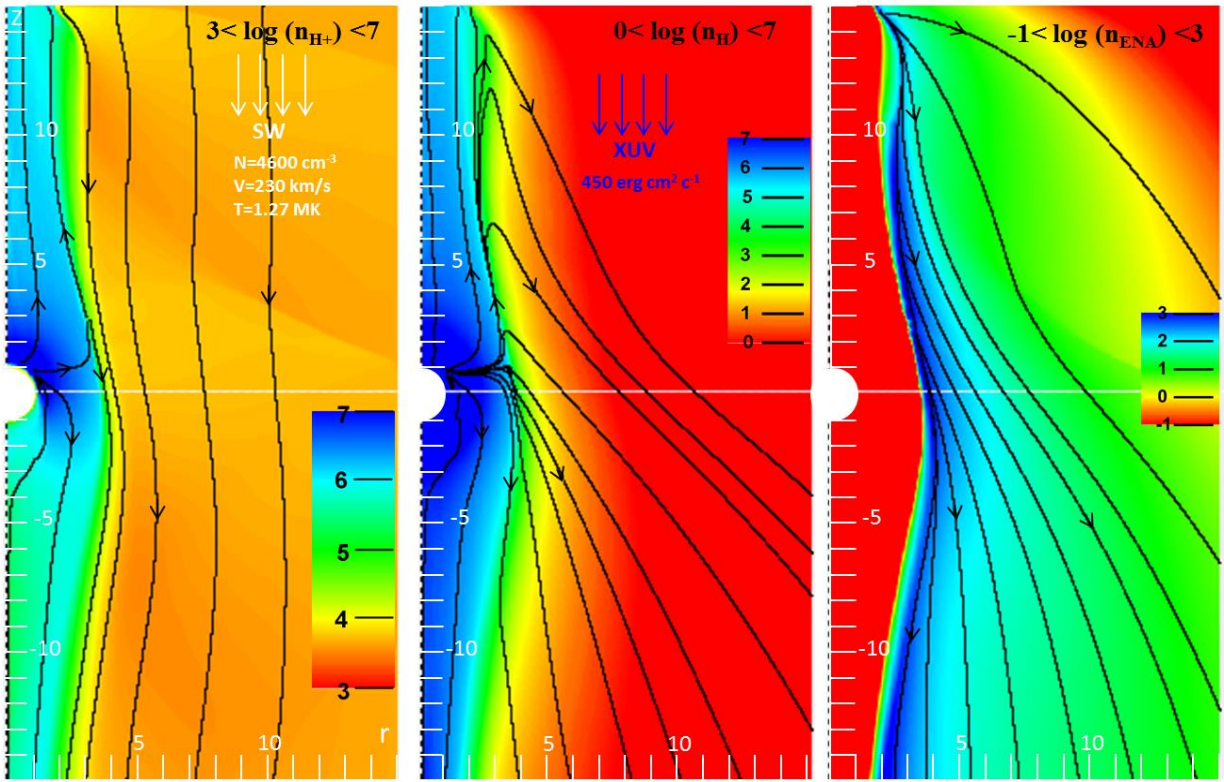

**Figure 5.** Density distributions of protons ($n_{H+}$), planetary atoms ($n_H$) and ENAs ($n_{ENA}$), realized in the *"captured by the star"* regime of PW and SW interaction around HD209458b for the total XUV flux at 1 a.u. $F_{XUV}=1$ erg s$^{-1}$ cm$^{-2}$ and under the conditions of *slow* SW with the total pressure $p_{sw} = 5 \cdot 10^{-6}$ μbar.

Figure 5 shows the density plots similar to ones in Figure 1, but calculated for the XUV flux of 1 erg cm$^{-2}$ s$^{-1}$. Qualitatively, these plots are very much the same and show that the ENAs are produced close to the PW stream boundary, penetrated by just a small number of atoms. The only difference between the Figures 1 and 5 consists in the particular amount of the generated ENAs, which could be seen in the figures color coding bars.

**3.3 The ENAs production mechanisms: radiation pressure versus charge-exchange**

In this section we compare the efficiency of two possible mechanisms for the production of ENAs in the close vicinity of a hot Jupiter, one of these mechanisms is related with the acceleration of neutral hydrogen atoms by the radiation pressure force $f_{rad} = F_{Ly\alpha,\lambda} \cdot \sigma_{Ly\alpha,\lambda}/c$ (*Vidal-Madjar et al. 2003, Lecavelier des Etangs et al. 2004, Lecavelier des Etangs et al. 2008*), and another, is due to the reaction of charge-exchange between the planetary atmospheric neutrals and fast protons of the SW (*Holmström et al. 2008, Ekenbäck et al. 2010, Kislyakova et al. 2014*). To quantify (distinguish) the effect of the radiation pressure, we performed the simulations with the radiation pressure, switched on and off in the presence and in the absence of a typical *slow* SW, taking the XUV flux at the lowest considered level of 1 erg cm$^{-2}$ s$^{-1}$ at 1 a.u. (i.e., 450 erg cm$^{-2}$ s$^{-1}$ at the planet), while the Lyα flux was kept the same as in previous simulations: $F_{Ly\alpha,\lambda} = 8\cdot 10^3$ erg cm$^{-2}$s$^{-1}$A$^{-1}$.

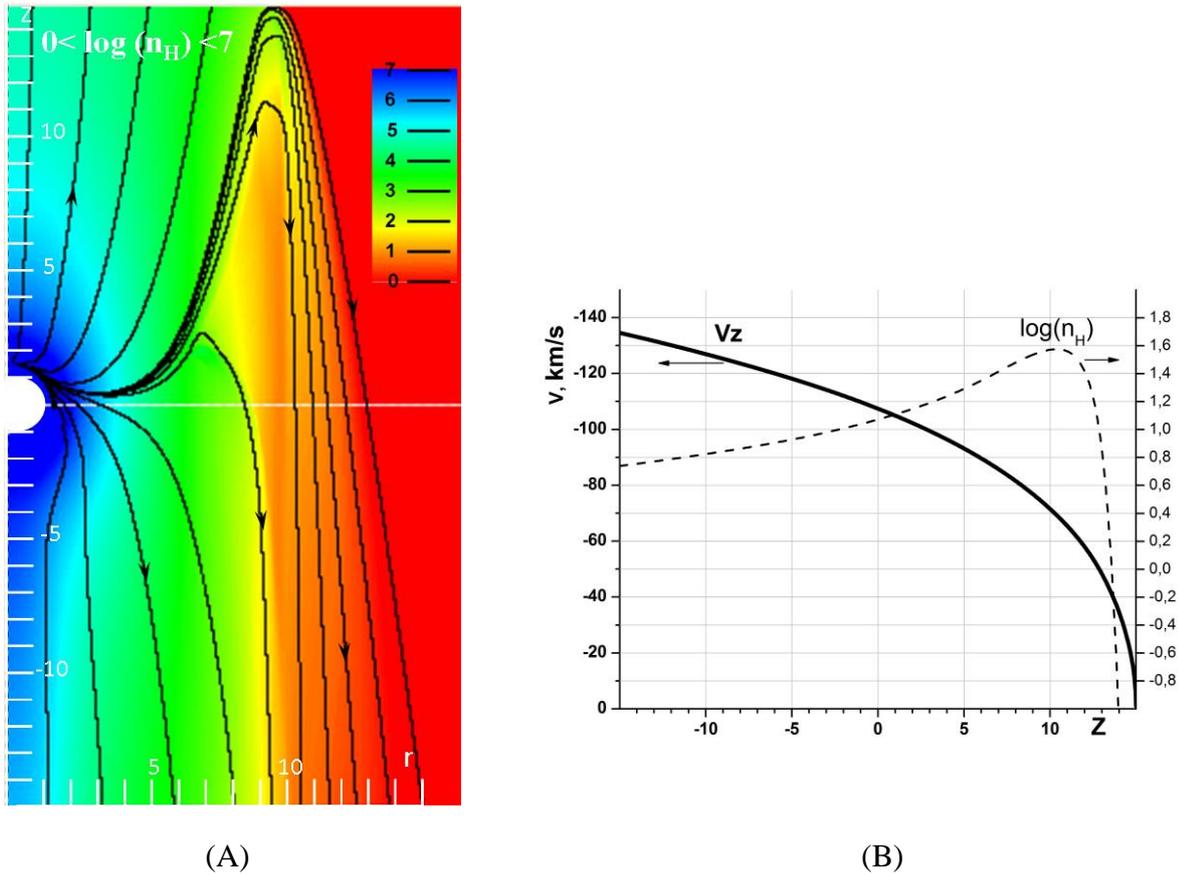

(A)  (B)

**Figure 6.**
(A): Density distribution of hydrogen atoms and their streamlines (black) simulated for HD209458b in the absence of SW, for the stellar Lyα flux at the line core $F_{Ly\alpha,\lambda} = 8\cdot 10^3$ erg cm$^{-2}$s$^{-1}$A$^{-1}$ and XUV flux $F_{XUV}$ = 450 erg cm$^{-2}$ s$^{-1}$ at the planet orbit (1 erg cm$^{-2}$ s$^{-1}$ at 1 a.u.).
(B): Profiles of the line of sight hydrogen velocity (solid) and density (dashed) along the planet-star direction at a fixed radial coordinate r = 11 R$_p$ under the same model parameters as in the case (A).

The plot in Figure 6A shows the modelling result in the case without SW. It is similar to one shown in Figures 1 and 5 with the typical day-side and night-side PW streams, having however a larger width $(7 \div 9)R_p$, which in the absence of SW is controlled by the competing processes of lateral thermal expansion and tidal acceleration. Such structure of the tidally locked hot Jupiter's PW without taking SW into account, has been obtained first in *Khodachenko et al.* (*2015*). Both, day- and nigh-side PW streams have sharp boundaries. However, some of the hydrogen atoms, passing close to these boundaries where density is rarified, are decoupled from protons and accelerated by the radiation pressure away from the star. The increase of their velocity is shown in Figure 6B. One can see that on the path of $20R_p$ the radiation pressure accelerates the hydrogen atoms up to velocity of more than 100 km/s, that is required for the interpretation of the HD209458b spectral transit features, supposing them to be caused by the Lyα *resonant thermal line broadening* absorption in the planetary ENA corona. However, the acceleration of atoms by the radiation pressure takes place only outside the relatively dense planetary stream, while inside the stream it is negligible partially due to the self-shielding of Lyα photons and mostly due to sharing of propulsion momentum received by atoms with much more numerous protons.

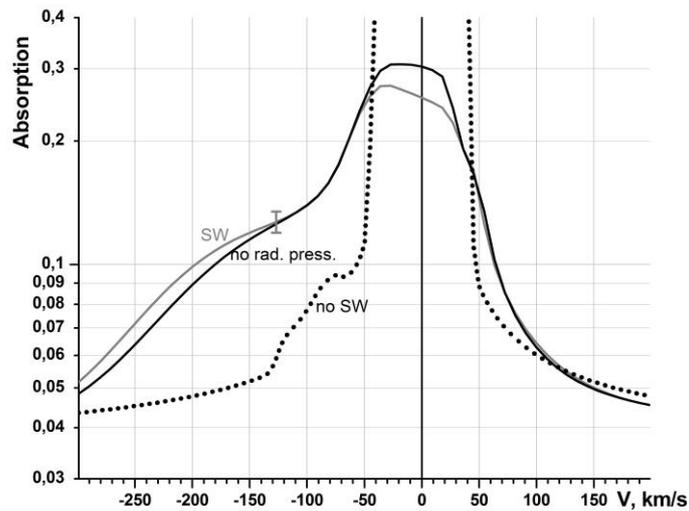

**Figure 7.** The profile of the Lyα absorption line for the *slow* SW is shown in the cases with the radiation pressure (grey solid) and without it (black solid). For comparison the case with no SW but with radiation pressure (dotted) is shown (same conditions as in Figure 6). XUV flux for all cases is 1 erg cm$^{-2}$ s$^{-1}$ at 1 a.u. An error bar shows the range of variations of absorption caused by the local density and velocity fluctuations.

The resulting Lyα absorption lines for the considered cases, i.e. with and without SW, as well as with the radiation pressure switched on and off, are presented in Figure 7. The effect of the radiation pressure is a clearly visible in the blue wing at the expected range of velocities up to 150 km/s in both cases, with and without SW. However, its total value is quite small and for the higher realistic XUV fluxes $F_{XUV} = 1 - 8$ erg cm$^{-2}$ s$^{-1}$ at 1 a.u. it is negligible.

The smallness of the radiation pressure effect is additionally demonstrated in the same figure by (showing) the calculation (results) performed under the typical *slow* SW conditions with and without account of the radiation pressure. One can see that the difference between these cases doesn't exceed 1%, and is within natural time variability of absorption caused by the local density and velocity fluctuations due to the instabilities developing at the contact boundary between PW and SW. Therefore, the performed comparison of the ENAs generation mechanisms reveals that under the considered parameters of the hot Jupiter HD209458b, the production of ENAs due to the atoms acceleration by Lyα radiation pressure appears insignificant. The ENAs, in the considered case (if they are created), are mainly produced by the reaction of charge-exchange between the planetary atoms and SW protons. Thus, the observed absorption features during the planet transits are caused by the *non-resonant natural line broadening*.

In spite of the fact that the radiation pressure is in principle able to accelerate the escaping planetary hydrogen atoms up to the velocities typical for the observed transit spectral features (see in Figure 6B), our hydrodynamic model shows that under the conditions of HD209458b, the amount of such accelerated particles is not sufficient to produce the measurable effects in the Lyα absorption line at the velocities of the order of –100 km/s. According to the simulations, only a small number of atoms can reach the rarified region where they may be efficiently accelerated by the radiation pressure. Even if to assume, that due to more complex 3D structure of the PW streams the Lyα radiation can penetrate into their denser layers, as argued in *Shaikhislamov et al. (2016)*, its acceleration effectiveness will be drastically reduced by the coupling between hydrogen atoms and protons via the resonant charge-exchange. It can be estimated analytically, that under the typical conditions expected at HD209458b the effect of acceleration by the radiation pressure force $f_{rad} = F_{Ly\alpha,\lambda} \cdot \sigma_{Ly\alpha,\lambda}/c$ is much weaker in the production of ENAs, as compared to the charge-exchange. The latter can be represented as an acceleration of an atom to a velocity of stellar proton $V_{SW}$ with a rate, equal to the charge-exchange reaction rate $\Gamma_{exch} = n_{SW} V_{SW} \sigma_{exch}$ which may be attributed to an effective volume force $f_{exch}$. The ratio of the radiation pressure and the effective charge-exchange forces then looks as follows:

$$\frac{f_{rad}}{f_{exch}} = \frac{\sigma_{Ly\alpha,\lambda} \cdot F_{Ly\alpha,0}}{\sigma_{exch} \cdot n_{SW} m_p V_{SW}^2 c} = \frac{\sigma_{Ly\alpha}}{\sigma_{exch}} \cdot \frac{P_{rad}}{P_{SW}} \quad (3)$$

Here $\sigma_{Ly\alpha,\lambda} = 5.5 \cdot 10^{-15} cm^2 \cdot A$, $P_{rad} = c^{-1} \cdot \int F_{Ly\alpha,\lambda} d\lambda$, $\sigma_{Ly\alpha} = \sigma_{Ly\alpha,\lambda}/\Delta\lambda$, $\Delta\lambda = \frac{1}{F_{Ly\alpha,0}} \int F_{Ly\alpha,\lambda} d\lambda$ – the width of the radiated stellar Lyα line which is about $0.75\,\text{Å}$. The typical SW pressure at the considered distance is about $P_{SW} \sim 5 \cdot 10^{-6}\,\mu bar$, while the radiation pressure of the Lyα flux $F_{Ly\alpha,0} = 8 \cdot 10^3\,erg\,cm^{-2} s^{-1} \text{Å}^{-1}$ at the line center is about 25 times smaller $P_{rad} \approx 2 \cdot 10^{-7}\,\mu bar$. At typical $V_{SW} \sim 300\,km/s$ the charge-exchange cross-section is $\sigma_{exch}(v) \sim 2 \cdot 10^{-15} cm^2$. So, finally we obtain $f_{rad}/f_{exch} \sim 0.15$. Therefore, even in spite of the fact that the stellar radiation force for the HD209458 is several times larger than its gravity, it is nevertheless significantly smaller than the effective "acceleration" of atoms due to SW. These analytical estimates are in full agreement with the results of our simulations.

**3.4 Lyα absorption at "*blown by the wind*" regime of PW and SW interaction**

It has been shown in *Shaikhislamov et al. (2016)*, that at the "*blown by the wind*" regime of the PW and SW interaction, when the SW is strong enough to overcome the stellar gravitation acting on the escaping planetary material, and to confine it within a kind of a paraboloid-shaped region bounded by an ionopause, more ENAs are generated. This is because the SW forms a bowshock, and planetary atoms can directly penetrate into it through the ionopause. It has been found that for HD209458b such regime realizes for a relatively strong SW, with the pressure of about 25 times larger than that of the typical *slow* SW considered above in this paper.

Figure 8 shows the results of calculation with the following parameters: XUV flux at 1 a.u. $F_{XUV}=4$ erg s$^{-1}$ cm$^{-2}$; $n_{sw} = 2.5 \cdot 10^4\,cm^{-3}$, $V_{sw} = 500\,km/s$, $T_{sw} = 2.90\,MK$ (i.e. the *fast* SW with

$p_{sw} = 1.3 \cdot 10^{-4} \mu bar$). One can see from the $n_{H+}$ plot that the sub-stellar position of the ionopause is at ~3$R_p$. This is below the first Lagrange point $R_{L1} \approx 4.1 R_p$. The PW flow is fully redirected towards the tail. The stellar protons form a bowshock and a shocked region, penetrated by planetary atoms. Differently from the "*captured by the star*" regime, in this case a significant amount of atoms, expanding from the day-side of the planet, reach the SW region. The ENAs are formed in a relatively thin layer near the ionopause, where the product of the SW protons and planetary atom densities has its maximum value. Practically all the ENAs are confined within the shocked region, i.e. between the ionopause and the bowshock.

Figure 9 shows the Lyα absorption profiles and their decompositions (done similarly as in Figure 2) for two sets of conditions, the first one (Figure 9A) is identical to that used in the simulations presented in Figure 8, whereas the second is characterized by the two times higher XUV flux and SW density. The the results show a distinct asymmetry of the absorption line in the case in Figure 9A, similar to the above considered cases of the low XUV fluxes (see in Figure 3). However, in the "*blown by the wind*" regime, a significant part of the absorption at the high velocity blue wing is produced by the relatively fast atoms in the transition layer of the ionopause. At the ionopause, the proton velocity and temperature rapidly change from the relatively low values of the expanding PW to the high values in the shocked SW. In such the transition layer the proton density is sufficiently high for strong coupling with atoms, resulting in their efficient pick-up due to the momentum exchange (*Shaikhislamov et al. 2016*).

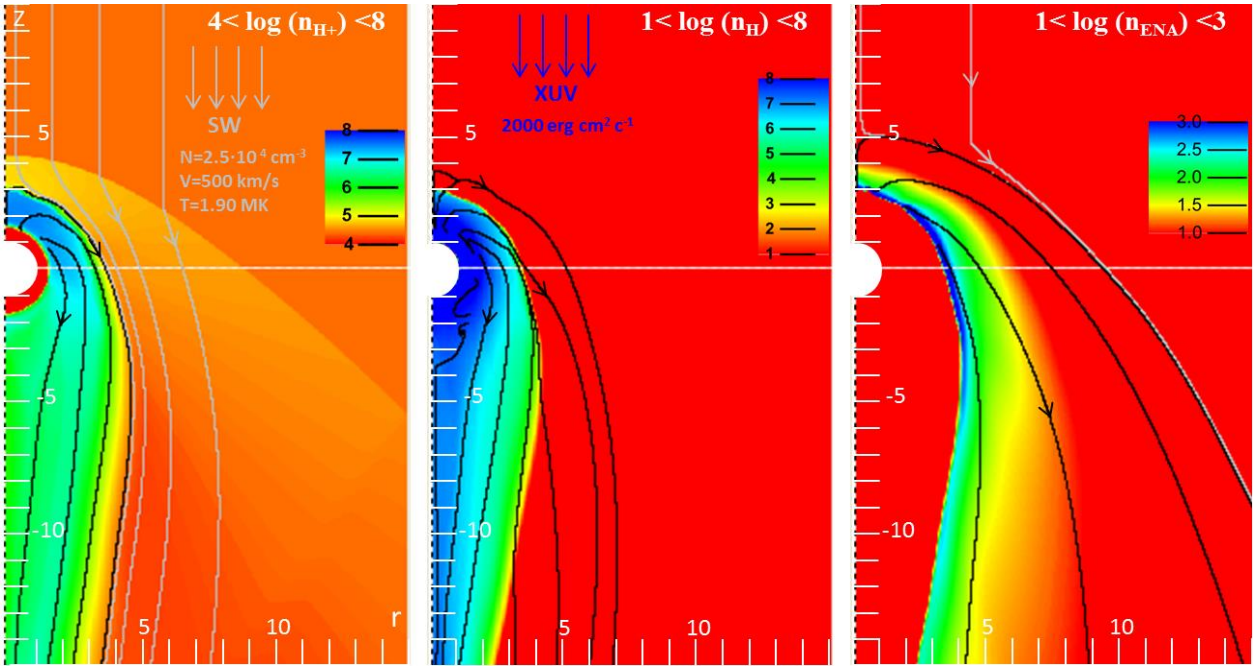

**Figure 8.** Density distributions of protons ($n_{H+}$), planetary atoms ($n_H$) and ENAs ($n_{ENA}$), realized in the "*blown by the wind*" regime of PW and SW interaction around HD209458b for the total XUV flux at 1 a.u. $F_{XUV}$=4 erg s$^{-1}$ cm$^{-2}$ and under the conditions of *fast* SW with the total pressure $p_{sw} = 1.3 \cdot 10^{-4} \mu bar$. With gray the stream lines of SW are shown.

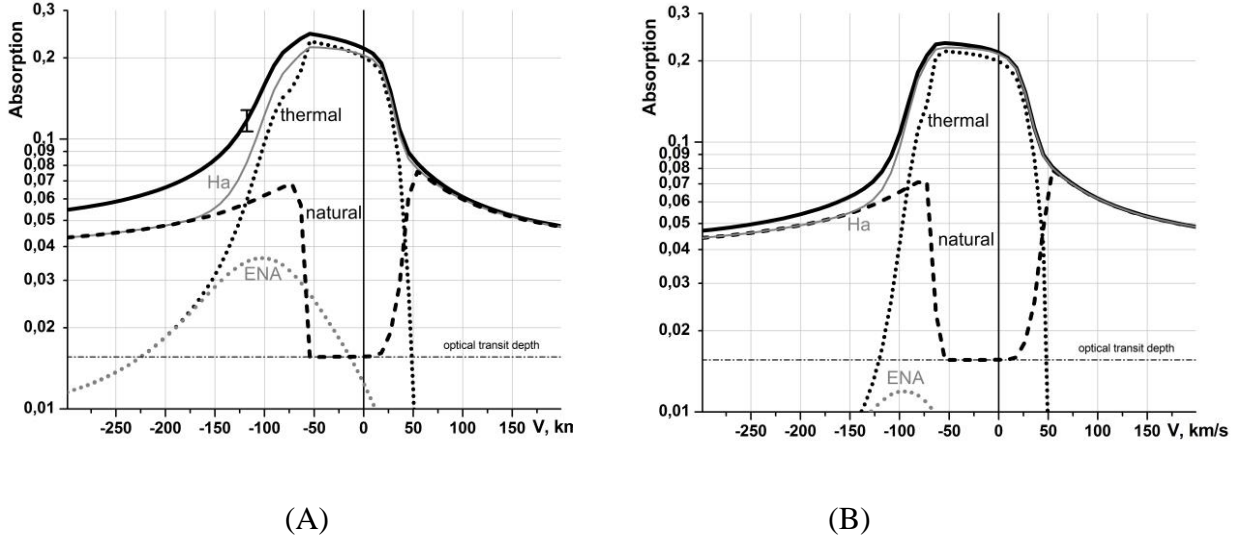

(A)                                      (B)

**Figure 9.** The profile of the Lyα absorption line in the *"blown by the wind"* regime of PW and SW interaction around HD209458b under the conditions of *fast* SW and different XUV fluxes. The decomposition of the total absorption line is marked similarly as in Figure 2. The gray solid line marked with Ha shows the absorption caused by planetary atomic hydrogen.

<u>(A)</u>:      $F_{XUV} = 4\,\text{erg cm}^{-2}\text{s}^{-1}$ at 1 a.u.,     $n_{sw} = 2.5 \cdot 10^4\,\text{cm}^{-3}$,     $V_{sw} = 500\,\text{km/s}$, $T_{sw} = 2.90\,\text{MK}$ (like in Figure 8).

<u>(B)</u>:      $F_{XUV} = 8\,\text{erg cm}^{-2}\text{s}^{-1}$ at 1 a.u.,     $n_{sw} = 5 \cdot 10^4\,\text{cm}^{-3}$,     $V_{sw} = 500\,\text{km/s}$, $T_{sw} = 2.90\,\text{MK}$.

This scenario has been tested with the simulations under the same conditions as in Figure 8, but with the terms, responsible for the production of ENAs due to charge-exchange and the radiation pressure, switched-off. At higher XUV flux (Figure 9B) the role of ENAs, produced by charge-exchange, becomes totally negligible and the asymmetry producing additional absorption at the blue wing is related with the pick-up of atoms in the transition layer of ionopause.

It has been also confirmed with the dedicated simulation runs, that the contribution of the radiation pressure to the production of ENAs is small under the considered system parameters expected for the HD209458b, and it increases the a Lyα absorption by less than just 1%.

## 4. Discussion and conclusions

In the present paper we simulated the Lyα absorption line at Doppler shifted velocities of the order of 100 km/s for the transits of HD209458b using for the first time the self-consistent 2D hydrodynamic model of the expanding PW which interacts with the plasma flow of SW in the vicinity of the planet. The model enables the detailed quantitative description of the structure of the ENA envelope around the planet, formed due to charge-exchange between the planetary hydrogen and stellar protons, as well as by the acceleration of atoms by the radiation pressure. It has been demonstrated that the most crucial factor affecting the ENA environment of the planet is the XUV flux. At the same time, it was shown that for typical XUV fluxes expected at HD209458 during a moderate or high level of activity, e.g., $4 \div 16\,\text{erg cm}^{-2}\text{s}^{-1}$ at 1 a.u., and for moderate SW, the corresponding Lyα absorption appears to be symmetric, resulting in the transit depth at the level of $6 \div 6.3\%$ which is equally produced in the Lyα blue and red wings by the *non-resonant natural line*

*broadening* absorption, taking place in the warm expanding upper atmosphere of the HD209458b deep inside of the Roche lobe. By this, the contribution of ENAs in the absorption process under the considered conditions remains negligible. This result is in full agreement with the conclusions made by *Ben-Jaffel* (*2007, 2008*) and *Koskinen et al.* (*2010*) who used the 1D models of the expanding exosphere of HD209458b. The novel feature found in the present work, consists of the fact that the intensity of PW and the integral mass loss cannot be inferred from this kind of absorption. With a series of dedicated modelling runs, the Lyα *non-resonant natural line broadening* absorption has been shown to be insensitive to the level of XUV flux and the intensity of SW in a wide range of values, which at the same time, to significant degree define the planetary upper atmosphere escape and the related mass loss.

Under the typical conditions, expected on HD209458b, the Lyα *resonant*, or *thermal line broadening* absorption related with the fast hydrogen atoms (ENAs), either accelerated by the radiation pressure or produced during charge-exchange with the SW protons, is small. However, under certain (somewhat unusual, but nevertheless possible) parameters of the SW and stellar radiation, the ENAs population increases. This finally results in the increase of the *thermal line broadening* absorption. In particular, it has been found, that at XUV flux at 1 a.u. below $2\,\mathrm{erg\,cm^{-2}\,s^{-1}}$, the increase of amount of ENAs results in perceptible increase of absorption at the blue wing of Lyα line. Under the typical SW parameters, the Lyα profile becomes distinctly asymmetric with the absorption level in the blue wing increased by several percent. The main reason for this is that at lower XUV fluxes the planetary hydrogen atoms survive longer time before being photo-ionized. This increases the amount of atoms that reach the boundary of PW and SW interaction region and penetrate into the SW where they undergo the charge-exchange and form ENAs.

Detailed analysis shows that in the relatively rarified exosphere, which is optically thin to XUV photons, about 95% of ionization happens due to the VUV photons in 500-912 Å range. The ionization by EUV and X-ray photons takes place in a dense optically thick upper atmosphere where VUV photons do not penetrate. In the spectra, used for our modelling the energy contained in the 500-912 Å range constitutes 22% of the total energy in the whole considered range of 20-912 Å. Thus, the total XUV flux at 1 a.u. of 2 erg cm$^{-2}$ s$^{-1}$, considered in the model, includes 0.44 erg cm$^{-2}$ s$^{-1}$ of its VUV part. While no measurements in VUV can be made in other stars, the Sun may be considered as a kind of proxy in that respect. The data regarding the solar spectral luminosity (e.g., *Tobiska 1993*, *Ribas et al. 2005*, *Linsky et al. 2013*) and the related coronal modeling reveal, in particular, that variation of VUV flux between solar maximum and minimum is about 2 times (for EUV it can be up to 3 times). At the same time, during the activity minimum, a kind of a minimal flux is achieved, which level is not related with any magnetic activity and is determined only by the basic stellar parameters, such as age, mass and temperature (i.e. the stellar type). The solar irradiance at 1 a.u., measured during the last activity minimum in 2008, was of about 1.9 erg cm$^{-2}$ s$^{-1}$ and 0.46 erg cm$^{-2}$ s$^{-1}$ in the ranges 100-912 Å and 500-912 Å, respectively (see data analysis by *Linsky et al. 2013*). Close values have been also measured at 1 a.u. in the solar minimum of 1974, e.g., 2.1 erg cm$^{-2}$ s$^{-1}$ and 0.42 erg cm$^{-2}$ s$^{-1}$ in the ranges 100-912 Å and 500-912 Å, respectively (*Heroux and Hinteregger 1978*). These values are close to those, at which the effect of ENAs becomes noticeable in our model. The star HD209458 is rather similar to the Sun, so use of the solar values as a kind of reference, seems reasonable. However, because of slightly larger size and total luminosity of HD209458, as compared to the Sun, some particular figures might be up to be 30-50% higher. The reconstruction of ionizing flux (i.e., <912 Å) of HD209458, based on the near UV measurements with HST COS instrument and X-ray measurements on XMM-Newton, using the Differential Emission Measure coronal models, gave the values at 1 a.u. of 6.4 erg cm$^{-2}$ s$^{-1}$ (*Louden et al. 2017*) and 2 erg cm$^{-2}$ s$^{-1}$ (*Sanz-Forcada et al. 2011*). Such a difference in the estimations of ionizing flux comes from the uncertainty of X-ray measurements. At the same time, it could be concluded, that the possibility of the XUV radiation flux of HD209458 at 1 a.u. at the level of 2 erg cm$^{-2}$ s$^{-1}$, or even lower, cannot be excluded. It should be also noted that, so far no direct or indirect data on the VUV for the stars is available, whereas for the Sun the measurements remain rather fragmented and does not cover many solar cycles. Nevertheless, a bit higher flux in the range 4÷8 erg cm$^{-2}$ s$^{-1}$ at 1 a.u. seems to be more realistic. Altogether the above arguments might support the idea of a possibly lower level of the ionizing radiation flux on the HD209458, which provides the conditions under which a relatively large number of the upper atmospheric hydrogen atoms in HD209458b reaches the ionopause boundary and penetrates into the SW region, where it produces by charge-exchange the sufficient amount of ENAs capable to result in the measured Lyα absorption profiles during the planetary transits.

There is another aspect, which concerns the atomic hydrogen content in the expanding upper atmosphere of a hot Jupiter. As it has been pointed out in our previous paper (*Shaikhislamov et al. 2016*), the comparison of different models dedicated to the study of PW generation reveals the significant diversity in the predicted estimates of radius of half-ionization (the distance from planet where $n_H = n_{H+}$) for HD209458b. It varies from 1.6$R_p$ in *Murray Clay et al.* (*2009*) up to 3.1$R_p$ in *Koskinen et al.* (*2013*) and 4$R_p$ in *García Muñoz* (*2007*). This difference comes from either specific stellar XUV spectrum used in the models, or boundary conditions, as well as different atmospheric composition and differently involved photo-chemistry. In our present calculations, the half-ionization radius measured in the sub-stellar direction varies from 2.5$R_p$ to 4.2$R_p$ for the XUV flux changing from 4 erg cm$^{-2}$ s$^{-1}$ to 1 erg cm$^{-2}$ s$^{-1}$, respectively. The first our figure is significantly less, whereas the second one is practically equal to the value obtained in *García Muñoz (2007)* under much higher XUV flux of 6.4 erg cm$^{-2}$ s$^{-1}$.

The difference in the atomic hydrogen content results in different values for the Lyα *non-resonant natural line broadening* absorption, e.g., 6.3% in our simulations, 6.6% in *Koskinen et al. (2010)* and 8%, calculated by *Ben-Jaffel (2008)* based on the model by *García Muñoz (2007).* At the same time, all these results does not differ more than an error and variability margins. While the value of 8% is most close to the absorption measured in high resolution observations (*Ben-Jaffel 2007*), it is not consistent with the presence of even a moderate SW. Otherwise, as it can be judged from our modeling, the expected large content of planetary hydrogen atoms should produce enough ENAs via charge-exchange, which will be visible as a strong absorption in the blue wing. Such contradictions and discrepancies leave the room for the improving the models of the hot Jupiters' upper thermospheres and exospheres on the one hand, and show the necessity of further spectral measurements of the planetary transits on the other hand.

Besides of the low XUV flux, we found other conditions, at which the observed additional absorption at the blue wing of Lyα can take place. These are realized in the case of a strong SW typical for strong Solar CME. In this case, as it was also predicted in our previous paper (*Shaikhislamov et al. 2016*), the *"blown by the wind"* regime of PW and SW interaction is realized, when the SW stops the escaping planetary material sweeping it towards the tail, and more ENAs are produced, as compared to the case of the *"captured by the star"* regime, when the material outflowing from the dayside of the planet flies in the direction of star. To exceed the tidal pull in the case of HD209458b, the SW ram pressure has to be an order of magnitude higher than its typical value in the quiet solar wind. In particular, for the SW pressures above $\sim 10^{-4}$ μbar at orbital distance of 0.047 a.u., under the "*blown by the wind*" regime of the PW and SW interaction the amount of atoms penetrating through the ionopause into the shocked region, formed around the planetary obstacle by the SW and filled by its fast and hot protons, greatly increases. That results in more effective production of ENAs which amount becomes large enough to explain the observations even at XUV fluxes at 1 a.u. higher than 4 erg cm$^{-2}$ s$^{-1}$. Note that the ENAs are formed in this case not only by charge-exchange between the planetary hydrogen atoms and protons, but also via momentum exchange and direct pick-up of atoms by protons in the boundary layer.

In all the explored possible conditions expected at HD209458b, including also the case of no SW, we found that the contribution of the radiation pressure (as an ENA accelerating factor) to the formation of Lyα absorption line doesn't exceed the value of 0.01, and in the most cases is even less than that. At the current level of sensitivity and reproducibility of the measurements it cannot be distinguished relative the symmetric *non-resonant natural line broadening* absorption at the level beyond 6%. Moreover, analytical estimations, supported by simulations, show that for the typical Sun-like SW, the action of the SW total pressure and charge-exchange produce a significantly stronger effect than that of the Lyα radiation pressure.

The performed study was done under the assumption of a non- or weakly- magnetized planet with a small intrinsic planetary magnetic dipole moment, which according to the planetary dynamo scaling laws is a reasonable approximation in the case of HD09458b (*Grießmeier et al. 2004, 2005, Khodachenko et al. 2012*). It should be noted however, that for a magnetized analog of HD209458b (*Trammell et al. 2014, Khodachenko et al. 2015*), the planetary magnetic field influences the character of material escape and results in appearance of a dead zone where the plasma is stagnant and a wind zone of the outflowing plasma. According *Khodachenko et al.* (*2015*), a relatively strong planetary magnetic field > 0.3G at the equator causes sufficiently large size of the dead zone, which reduces the overall material escape from the planet. This in its turn should affect the interaction of the escaping PW with SW and the production of ENAs. In view of the strong coupling between protons and neutrals in the PW, in the case of a large dead zone, less planetary atoms will be able to reach and penetrate the stellar wind, and therefore less ENAs will be finally produced.

While the present study does not give essentially new interpretation of HD209458b transit spectral observations, the analysis of various mechanisms of Lyα absorption is revealing and can be useful for other systems, as well as for the diagnostics of the dynamical plasma environments in the vicinity of the planet and its mass loss. Among other exoplanets in that respect, the most promising seems to be the warm Neptune GJ 436b. Its small size relative to the stellar disk makes the *non-resonant natural line broadening* absorption insignificant, while the low XUV flux and extended weakly bounded by the planet's gravity exosphere, provide favorable conditions for the generation of ENAs by charge-exchange.

### Acknowledgements

This work was supported by grant № 16-52-14006 of the Russian Fund of Basic Research, RAS SB research program (project II.10 №0307-2016-0002), as well as by the projects P25587-N27, S11606-N16 and S11607-N16 of the Austrian Science Foundation (FWF). MLK also acknowledges the FWF projects I2939-N27, P25640-N27, Leverhulme Trust grant IN-2014-016 and RME&S – Europlanet2020-RI collaboration grant №14.616.21.0084. Parallel computing simulations, key for this study, have been performed at Computation Center of Novosibirsk State University, SB RAS Siberian Supercomputer Center, and Supercomputing Center of the Lomonosov Moscow State University.

### References

Ben-Jaffel, L., 2007, ApJ, 671, L61.

Ben-Jaffel, L., 2008, ApJ, 688, 1352.

Ben-Jaffel, L., Sona Hosseini, S., 2010, ApJ, 709(2), 1284.

Bisikalo, D., Kaygorodov, P., Ionov, D., et al. 2013, ApJ, 764, art.id. 19.

Bourrier, V., Lecavelier des Etangs, A., 2013, A&A, 557, id.A124.

Bourrier, V., Lecavelier des Etangs, A., Dupuy, H., et al., 2013, A&A, 551, id.A63

Bourrier, V., Ehrenreich, D., Lecavelier des Etangs, A., 2015, A&A, 582, id.A65.

Christie, D., Arras, P., Li, Z.-Y., 2016, ApJ, 820, art.id.3.

Ehrenreich, D., Des Etangs, A. L., Hébrard, G., et al., 2008, A&A, 483, 933.

Ehrenreich, D., Bourrier, V., Bonfils, X., et al., 2012 A&A, 547, id.A18.

Ehrenreich, D., Bourrier, V., Wheatley, P.J., et al., 2015, Nature, 522, 459.

Ekenbäck, A., Holmström, M., Wurz, P., et al. 2010, ApJ, 709, 670.

Erkaev, N.V., Penz, T., Lammer, et al., ApJS, 2005, 157, 396-401.

Fossati, L., Haswell, C. A., Froning, C. S., et al., 2010, ApJL, 714, L222

García Muñoz, A. 2007, P&SS, 55, 1426.

Grießmeier, J.-M., Stadelmann, A., Motschmann, U., et al. 2005, Astrobiology, 5, 587

Grießmeier, J.-M., Stadelmann, A., Penz, T., et al. 2004, A&A, 425, 753

Guo, J. H. 2011, ApJ, 733, 98.

Heroux L., Hinteregger H. E., 1978, J. Geophys. Res.: Space Physics, 83, A11, 5305-5308

Holmström, M., Ekenbäck, A., Selsis, F., et al. 2008, Nature, 451, 970.

Johnstone, C. P., Güdel, M., Lüftinger, T., et al., 2015, A&A, 577, A27.

Khodachenko, M. L., Lammer, H., Lichtenegger, H. I. M., et al. 2007a, Planet. Space Sci., 55, 631.

Khodachenko, M. L., Ribas, I., Lammer, H., et al. 2007b, Astrobiology, 7, 167.

Khodachenko, M. L., Alexeev, I. I., Belenkaya, E., et al. 2012, ApJ, 744, 70

Khodachenko, M.L., Shaikhislamov, I.F., Lammer, H., et al., 2015, ApJ, 813, art.id.50.

Kislyakova, G. K., Holmström, M, Lammer, H., et al., 2014, Science, 346, 981.

Koskinen, T. T., Yelle, R. V., Lavvas, P., & Lewis, N. K., 2010. ApJ, 723, 116.

Koskinen, T. T., Harris, M. J., Yelle, R. V., et al., 2013, Icarus, 226, 1678.

Lammer, H., Selsis, F., Ribas, I., et al., 2003, ApJ, 598, L121.

Lammer, H., Odert, P., Leitzinger, et al., 2009, A&A, 506(1), 399-410.

Lecavelier des Etangs, A., Vidal-Madjar, A., McConnell, J. C., et al., 2004, A&A, 418, L1.

Lecavelier des Etangs,A., Vidal-Madjar, A., Desert, J.-M., 2008, Nature, 456, E1.

Lecavelier des Etangs, A., Ehrenreich, D., Vidal-Madjar, A., et al. 2010, A&A, 514, A72

Linsky, J.L., Yang, H., France, K., et al. 2010, ApJ, 717, 1291.

Linsky, J. L., Fontenla, J., & France, K., 2013, ApJ, 780(1), 61.

Louden, T., Wheatley, P. J., & Briggs, K., 2017, MNRAS, 464(2), 2396-2402.

Matsakos, T., Uribe, A., Königl, A., 2015, A&A, 578, A6.

Murray-Clay, R. A., Chiang, E. I., Murray, N., 2009, ApJ, 693, 23.

Penz, T., Erkaev, N.V., Kulikov, Yu. N., et al., 2005, ApJ, 622(1), 680.

Ribas, I., Guinan, E. F., Güdel, M., Audard, M., 2005, ApJ, 622, 680

Sanz-Forcada, J., Micela, G., Ribas, et al., 2011, Astronomy & Astrophysics, 532, A6.

Shaikhislamov, I. F., Khodachenko, M. L., Sasunov, Yu. L., et al., 2014, ApJ, 795, art.id. 132.

Shaikhislamov I. F., Khodachenko M. L., Lammer H., et al., 2016, ApJ, 832, N2, 173

Stone, J. M., Proga, D., 2009, ApJ, 694, 205.

Tasitsiomi, A., 2006, ApJ, 645(2), 792.

Tian, F., Toon, O. B., Pavlov, A. A. & De Sterck, H., 2005, ApJ, 621, 1049.

Tobiska W. K., 1993, JGR: Space Physics, 98(A11), 18, 879.

Trammell, G. B., Arras, .P & Li, Z.-Y., 2011, ApJ, 728, 152.

Trammell, G. B., Li, Z. Y., & Arras, P. 2014, ApJ, 788, 161

Tremblin, P., Chiang, E., 2013, MNRAS, 428, 2565.

Tripathi, A., Kratter, K. M., Murray-Clay, R. A., & Krumholz, M. R., 2015, ApJ, 808(2), 173.

Vidal-Madjar, A., Lecavelier des Etangs, A., Désert, J.M., et al., 2003, Nature, 422, 143.

Vidal-Madjar, A., Désert, J., Lecavelier des Etangs, A., et al., 2004, ApJ, 604, L69.

Vidal-Madjar, A., Lecavelier des Etangs, A., Désert, J.-M., et al., 2008, ApJ, 676, art.id. L57.

Wood, B. E., Redfield, S., Linsky, J. L., et al., 2005, ApJSS, 159(1), 118.

Yelle, R. V. 2004, Icarus, 170, 167.